\documentclass[twocolumn, superscriptaddress]{revtex4-2}

\usepackage[utf8]{inputenc}
\usepackage{graphicx}
\usepackage{amsmath,amssymb,mathtools}
\usepackage{bm}
\usepackage{dcolumn}
\usepackage{booktabs}
\usepackage{threeparttable}
\usepackage{tabularx}
\usepackage{wrapfig}
\usepackage{float}
\usepackage{stmaryrd}
\usepackage{xcolor}
\usepackage{nameref,hyperref}

\hypersetup{
    colorlinks=true,
    citecolor=red,
}

\newcommand{\total}{\mathrm{total}}
\newcommand{\diff}{\mathrm{diff}}
\newcommand{\direct}{\mathrm{direct}}

\newcommand{\uni}{Department of Computer Sciences, University of T\"ubingen, Germany}
\newcommand{\mpi}{Max Planck Institute for Biological Cybernetics, T\"ubingen, Germany}
\newcommand{\imp}{Department of Bioengineering, Imperial College London, London, UK}
\newcommand{\maa}{School of Business and Economics, Maastricht University, Maastricht, Netherlands}

\renewcommand{\th}{\mathrm{th}}
\newcommand{\rec}{\mathrm{rec}}
\newcommand{\inn}{\mathrm{in}}
\newcommand{\out}{\mathrm{out}} 
\newcommand{\layer}{\mathrm{layer}} 
\newcommand{\ms}{\mathrm{ms}} 
\newcommand{\adapt}{\mathrm{adapt}} 
\newcommand{\reg}{\mathrm{reg}} 
\newcommand{\target}{\mathrm{target}} 
\newcommand{\av}{\mathrm{av}} 
\newcommand{\Hz}{\mathrm{Hz}} 
\newcommand{\task}{\mathrm{task}} 

\begin{document}

\title{Diffusion of Neuromodulators for Temporal Credit Assignment}

\author{João Barretto-Bittar} 
\affiliation{\uni}
\affiliation{\mpi}
\affiliation{\maa}

\author{Anna Levina}
\affiliation{\uni}
\affiliation{\mpi}

\author{Emmanouil Giannakakis} 
\thanks{These authors contributed equally.}
\affiliation{\uni}
\affiliation{\mpi}
\affiliation{\imp}

\author{Roxana Zeraati}
\thanks{These authors contributed equally.}
\affiliation{\uni}
\affiliation{\mpi}

\begin{abstract}
Biological learning achieves temporal credit assignment despite sparse and imprecise feedback, often relying on neuromodulatory signals acting over space and time. Here, we introduce a learning mechanism in which error information diffuses locally through the network, similar to volume transmission of neuromodulators. This distributed modulation allows neurons to learn even in the absence of direct feedback, using the local concentration of the diffusing credit signal. Applied to recurrent spiking neural networks with sparse feedback connectivity, diffusive credit signaling improves learning across three benchmark tasks.  Using eligibility propagation as a baseline learning mechanism, we show how diffusion-based modulation can provide a plausible mechanism for credit assignment in sparsely connected neural circuits.
\end{abstract}
\maketitle

\title{Diffusion of Neuromodulators for Temporal Credit Assignment}

\author{Joao Barretto-Bittar} 
\affiliation{\uni}
\affiliation{\mpi}

\author{Emmanouil Giannakakis} 
\affiliation{\uni}
\affiliation{\mpi}

\author{Roxana Zeraati} 
\affiliation{\uni}
\affiliation{\mpi}

\author{Anna Levina}
\affiliation{\uni}
\affiliation{\mpi}

\maketitle

\section{Introduction}
Biological learning is a ubiquitous feature of living organisms. The nervous system of most animals is known to be highly adaptive, with a variety of local plasticity mechanisms and modulatory systems working in tight coordination to efficiently modify synaptic connectivity. Unlike biological networks, artificial neural networks (ANNs) are overwhelmingly trained via backpropagation of errors, a method of exact credit assignment that achieves very high performance in a wide range of tasks. The success of backpropagation for training ANNs has motivated numerous hypotheses that biological learning may be governed by similar principles \cite{lillicrap2020backpropagation}. However, several constraints of biological networks (non-exact credit assignment, sparse connectivity and feedback, etc.) make an exact implementation of standard backpropagation in biological networks unlikely, leading to a search for biologically plausible alternatives that could replicate the performance of backpropagation within the constraints of biological network connectivity and signalling mechanisms.

Eligibility propagation (e-prop) \cite{bellec2020solution} is among the most successful biologically plausible alternatives to backpropagation through time (BPTT). Its performance, however, deteriorates in networks with sparse feedback connectivity, architectures that more closely resemble biological network organization \cite{bullmore2012economy}.
Recent extensions incorporating neuromodulatory signals have achieved performance improvements by enriching the learning signal with additional structure or cell-type specific communication\cite{liu2021cell, liu2022biologically}. While effective, these approaches rely on precise and targeted credit assignment. In contrast, neuromodulatory systems predominantly operate through volume transmission, in which signals diffuse through the extracellular space and modulate populations of 
neurons over extended spatial scales \cite{cragg2004dancing, arbuthnott2007space, dayan2012twenty}.

Here, we examine a learning mechanism in which credit signals spread spatially through the network, with credit assignment determined by the local concentration of a modulatory particle rather than by its point of origin.

\section{Results}

\begin{figure}[t!]
\centering
\includegraphics[width=1\linewidth]{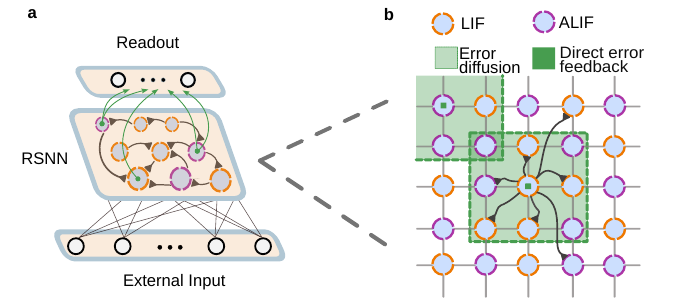}
\caption{Network architecture and local connectivity pattern. (\textit{a.}) A spatially embedded RSNN with local connectivity receives inputs from an external input layer. The activities of the RSSN are processed by a readout layer to generate the output. The RSNN is sparsely connected to the input and readout layers. (\textit{b.}) LIF and ALIF cells are randomly placed in an equally spaced 2D grid. Neurons are connected preferentially to nearby neurons. Only a few cells are connected to the readout layer and receive direct neuromodulatory feedback. Once upon arrival, the neuromodulatory signal diffuses to neighboring neurons.}
\label{fig:model}
\end{figure}

To assess the impact of diffusing credit signals on learning, we study recurrent spiking neural networks (RSNNs), learning to perform several complex temporal tasks. Each RSNN receives task-specific input as spike trains from an external input layer, and its activity is read out by an output layer composed of leaky non-spiking neurons (Fig.\ref{fig:model}a).

Our RSNNs consist of two neuron types: leaky integrate-and-fire (LIF) and its variant with firing-rate adaptation (ALIF). The proportion varies across tasks (more details in Supplementary Material). The neurons are randomly embedded on a uniformly spaced 2D grid, where the probability of a connection from neuron $i$ to neuron $j$ decreases exponentially with the squared distance between them (Fig. \ref{fig:model}b). The decay rate was set to achieve approximately 10\% connectivity. This arrangement promotes a local connectivity pattern, favoring links between nearby neurons. The connectivity to the input and readout layers is sparse, comprising a random 10\% subset of all possible connections, without bias toward either neuron type. Additional implementation details and model equations are provided in the Supplementary Material.

During each task, the RSNNs receive feedback credit signals that encode the network's task-related error, which modulates learning but not neuronal activity. Critically, we assume that these neuromodulatory signals do not operate with surgical precision. Rather, once released, they not only reach their target neurons but also diffuse through intercellular space to influence neighboring cells over several subsequent time steps (Fig. \ref{fig:model}b). 
Due to spatial diffusion, the total credit signal available to neuron $j$ at time step $t$, $C^{t, \total}_{j}$, is the sum of directly delivered credit signals, $C^{t, \direct}_j$, and those arriving via diffusion, $C^{t, \diff}_{j}$ 
\begin{equation}
C^{t, \total}_{j} = C^{t, \direct}_j + C^{t, \diff}_{j}.
\label{eq:totallearningsignal}
\end{equation}
The exact form of the direct credit signal is determined by the learning rule, for example, being proportional to the task loss in e-prop \cite{bellec2020solution}. Importantly, the feedback connectivity is also kept sparse (only 10\% of the possible connections). As a result, for most neurons, feedback credit signals are available only through diffused credit signals.

To capture biological processes that progressively reduce the availability of released neuromodulators, such as reuptake and enzymatic degradation, we assume that, at each time step, the local concentration of the neuromodulatory signal decays at a fixed rate $k \in [0,1]$ prior to diffusion. Diffusion redistributes the available neuromodulators between the original location and all the neighbours, resulting in the total neuromodulator arriving in the location $i$ to be
\begin{equation}
C^{t, \diff}_{j} = \sum_{i \in \mathcal{N}^M_{(j)}}{D_{ji} \cdot k \cdot C_i^{t-1, \total}}
\label{eq:diffusion}
\end{equation} 
 Here $D_{ji}$ denotes the fraction of the neuromodulatory signal available at location $i$ that is transferred to $j$ over one time step, and $\mathcal{N}^M_{(j)}$ defines the diffusion neighborhood of neuron $j$. By conservation law, we have that $\sum_{i \in \mathcal{N}^M_{(j)}}{D_{ji}} = 1$. In this work, we adopt the Moore neighborhood, which includes the neuron itself and its eight immediate neighbors. Consequently, a fraction of the available signal remains at neuron $j$ at each time step rather than diffusing away. We assume homogeneous diffusion within this neighborhood, such that the available signal is distributed uniformly across all neighbors, i.e., ($D_{ji}=\frac{1}{9}$). We efficiently simulate this diffusion-like mechanism  (Eq.~\ref{eq:diffusion}) using a Cellular Automaton (CA), enabling rapid computation of modulatory particle concentrations at every point in space. 

Our approach can be combined with any learning rule that incorporates feedback signals into its formulation. Here, we adopt eligibility propagation (e-prop) \cite{bellec2020solution}, which falls into this category and is one of the state-of-the-art biologically plausible learning rules for RSNNs.

E-prop's weight update rule factorizes into two distinct components: a local eligibility trace, $e^t_{ji}$, and a top-down modulatory learning signal $C^{t, \total}_{j}$, that represents the credit signal 
\begin{equation}
\Delta W_{ji} = \eta \sum_t{C^{t, \total}_je^t_{ji}},
\label{eq:e-prop}
\end{equation}
 with $\eta$ being the learning rate.  
 
 Those two terms are derived so that the update approximates Backpropagation Through Time (BPTT). Their exact expressions depend on the specific neuron and network models; for our network, they are provided in the Supplementary Material. In short, the eligibility trace acts as a decaying memory of pre- and postsynaptic neural activity, while the learning signal modulates the magnitude of the weight updates according to the network's error in the task.

\begin{figure*}[t]
\centering
\begin{minipage}[c]{0.6\textwidth}
    \centering
    \includegraphics[width=\linewidth]{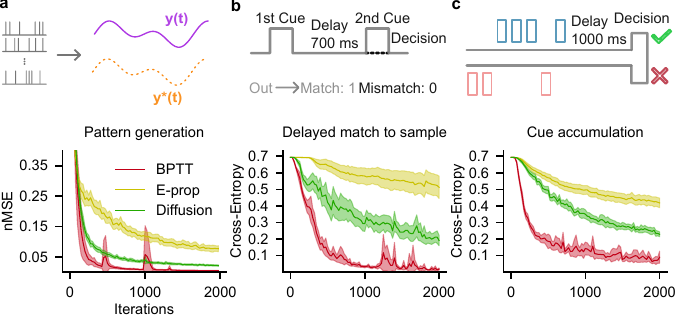}
\end{minipage}
\hfill
\begin{minipage}[c]{0.39\textwidth}
    \caption{Diffusion of neuromodulatory signal improves performance of e-prop when RSNN is sparsely connected to readout. \textit{Upper row}: task schematic. \textit{Lower}: learning curves for BPTT, e-prop, and e-prop with diffusion. Solid lines indicate the average value across 20 runs, and shaded areas the standard error of the mean. (\textit{a.}) Pattern generation, where for visualization purposes we plotted the normalized MSE $\text{ nMSE} = {\sum_{k,t} \left( y^*_{k,t} - y_{k,t} \right)^2}/{\sum_{k,t} \left( y^*_{k,t} \right)^2}$ for zero-mean target $y^*_{k,t}$. (\textit{b.}) Delayed match to sample. (\textit{c.}) Cue accumulation. The displayed results are for the post-release neuromodulatory signal decay rate of $k=0.75$. However, the results are qualitatively similar for $k \in \{ 0.25, 0.5, 0.9\}$, clearly outperforming the version without diffusion.}
    \label{fig:results}
\end{minipage}
\end{figure*}

Using e-prop with and without diffusion, we train our networks on three benchmark tasks: pattern generation, delayed match-to-sample, and cue accumulation \cite{liu2021cell}. In the first task, pattern generation (Fig.~\ref{fig:results}a), the network should learn to reproduce a 1-dimensional target signal composed of a weighted sum of five sinusoids, using realizations of Poisson noise as input. In this task, error feedback is provided at every time step. In contrast, the delayed match-to-sample and cue accumulation tasks provide error signals only at the final time frame, when the network must make a decision based on prior inputs. In the delayed match to sample task (Fig.~\ref{fig:results}b), the goal is to compare the values of two binary cues presented with a delay window between them, and then to determine whether the cues were identical (1-1 or 0-0) or different (1-0 or 0-1). Meanwhile, in the cue accumulation task (Fig.~\ref{fig:results}c), a sequence of seven cues is presented, each appearing on either the left or right side. Following a delay period with no cue, the network must indicate which side the majority of cues were displayed on.

We find that diffusing error signals significantly improves e-prop performance across all three tasks (Fig.~\ref{fig:results}) in the sparse-feedback connectivity setup. Compared with standard e-prop without diffusion, our variant consistently yields better learning outcomes, narrowing the performance gap with BPTT, which was included as a lower-bound comparison for the learning curves. 

Furthermore, although the local connectivity pattern of our RSNNs more closely reflects biological circuitry, we find that randomly connected sparse RSNNs benefit similarly from the local diffusion of credit signals in the tasks considered here.

\section{Discussion}

Temporal credit assignment with sparse feedback pathways is challenging, and even state-of-the-art biologically plausible rules, such as e-prop, struggle in this setting. While random e-prop \cite{bellec2020solution}, which employs random feedback weights, performs well in sparsely connected networks, it still presupposes dense feedback pathways in which each neuron receives its own dedicated error signal. In this regard, our work complements previous results \cite{liu2021cell}, which augmented e-prop with an additional cell-specific local neuromodulatory signal, also achieving a higher performance in settings with sparse feedback. However, that improvement relied on the direct communication of precise errors between connected neurons. Here, we show that a less precise form of neuromodulatory communication, which relies on chemical diffusion, can offer similar benefits for local learning.

Although it is well established that the brain relies on both synaptic and volume transmission for neuromodulation \cite{ozccete2024mechanisms}, the latter has received less attention in artificial neural networks. It has been suggested that such a mechanism can help mitigate catastrophic forgetting \cite{velez2017diffusion} and increase dynamical flexibility of the network by selectively modulating subsets of neurons \cite{tsuda2026neuromodulators}. Recent studies have also demonstrated that volume transmission of modulatory signals can enable gating properties and implement context factorization in RNNs \cite{costacurta2024structured, bullvolume}.

In addition to the encoding of prediction-error signals by dopamine \cite{glimcher2011understanding}, other neuromodulators influence biological learning, including serotonin, acetylcholine, and others \cite{brzosko2019neuromodulation}. 
We believe that our approach, due to its computational efficiency and inherent flexibility, provides a promising framework for exploring the functional roles of diffusing neuromodulators in artificial systems and for testing hypotheses about their biological counterparts.

Our results suggest that biochemical processes known to operate in biological circuits, such as the diffusion of modulatory substances, may play a functional role in enabling learning under realistic connectivity constraints. Our findings motivate further investigation into the interaction between neuromodulator dynamics and learning in biological systems and offer a direction for efficiently training spatially embedded artificial networks.

\section{Materials and Methods}
We trained sparsely connected recurrent spiking neural networks comprised of both LIF and ALIF units on the tasks described in \cite{liu2021cell}. For the training, we used eligibility propagation (e-prop) augmented with spatiotemporal diffusion of the credit signals. To emulate the diffusion, we employed a cellular Automaton. Further details are provided in the Supplementary Material

\section{Acknowledgements}
This work was supported by the German Federal Ministry of Education and Research (BMBF): Tübingen AI Center, FKZ: 01IS18039A (EG, RZ, AL), the Max Planck Society (RZ) and the Centre of Integrative Neuroscience at Maastricht University (JBB). We also thank the International Max Planck Research School for Intelligent Systems (IMPRS-IS) and the Joachim Herz Foundation for their support.

\bibliographystyle{unsrt}
\bibliography{bibliography}

\clearpage
\appendix
\onecolumngrid

\section{Network Model}
All simulations reported in this work are based on recurrent spiking neural networks (RSNNs) that receive spike-based inputs from an external input layer. An output layer reads out the activity of the RSNN and produces the network output.

\subsection{Spiking Neurons}
In line with \cite{bellec2020solution}, we adopt discrete-time recurrent spiking neural networks (RSNNs), consisting of a combination of leaky integrate-and-fire (LIF) and adaptive threshold LIF (ALIF) neurons \cite{teeter2018generalizedALIFneurons}. The overall network size and the proportion of LIF and ALIF neurons vary across tasks, with full architectural details provided in Section~\ref{sec:params}.  

The dynamics of the LIF and ALIF neurons in the recurrent layer are governed by the following equation for the membrane potential $v^t_j$
\begin{equation}
v^{t}_j = \alpha v^{t-1}_j + (1 - \alpha) (\sum_{i \neq j}{W^{\rec}_{ji}z^{t-1}_i} + \sum_{i}{W^{\inn}_{ji}x^{t}_i}) - z^{t-1}_j v_{\th},
\label{eq:v dynamics}
\end{equation}
with $\alpha=e^{\Delta t / \tau_m}$. Here $\Delta t$ is the simulation time step, fixed to $1\,\ms$ in all experiments, and $\tau_m \in \{20\,\ms$, $30\,\ms \}$ is the membrane time constant. 

We assume a transmission delay of $1 \ms$ for recurrent connections, such that recurrent spikes influence the postsynaptic membrane potential at the next time step, as reflected by the term $z^{t-1}_i$ in Eq.~\eqref{eq:v dynamics}. In contrast, external inputs are applied without delay and therefore enter the dynamics at the current time step via $x^t_i$.

$W^{\rec}_{ji}$ and $W^{\inn}_{ji}$ denote the synaptic weights from presynaptic neuron $i$ to postsynaptic neuron $j$, where neuron $i$ belongs to the recurrent layer or the external input layer, respectively. The variable $z_j^t$ indicates whether neuron $j$ emits a spike at time $t$ and is defined as
\begin{equation}
z^{t}_j = H\!\left(v^t_j - A^t_j\right),
\label{eq:spike_function}
\end{equation}
where $H(\cdot)$ is the Heaviside step function. 
Therefore, a spike $z^t_j$ is emitted whenever the membrane potential $v^t_j$ exceeds the neuron’s instantaneous firing threshold $A^t_j$, whose dynamics is given by
\begin{equation}
A^{t}_j = v_{\th} + \beta_j a^t_j,
\label{eq:adapt thr}
\end{equation}
where $v_{\th}$ is a baseline threshold and the variable $\beta_j$ controls the strength of spike-triggered adaptation by scaling the contribution of an adaptation variable $a^t_j$. The adaptation dynamics follows
\begin{equation}
a^{t}_j = \rho a^{t-1}_j + (1-\rho)z^{t-1}_j,
\label{eq:adapt dynamics}
\end{equation}
with $\rho = e^{\Delta t / \tau_{\adapt}}$ denoting the decay factor of the adaptation variable. The adaptation time constant $\tau_{\adapt}$ is chosen to match the temporal scale of the working memory demands of the task; for example, in the cue accumulation task, we set $\tau_{\adapt} = 2000\,\ms$.

For LIF neurons, the parameter $\beta_j$ in ~\ref{eq:adapt thr} is set to zero, effectively keeping the firing threshold fixed at its baseline value $v_{\th}$. Additionally, as in \cite{bellec2020solution, liu2021cell}, we implement a simple model of the refractory period where after neuron $j$ emits a spike, all state variables are allowed to evolve freely, except for the spike output $z^t_j$, which is clamped to zero for a short period of $2$--$5\,\ms$, depending on the task. A detailed description of the parameter choices across tasks is provided in Section~\ref{sec:params}.

\subsection{Network Output and Task Loss Function}
The network output $y^{t}_k$ is generated by leaky non-spiking readout neurons, with dynamics described by 
\begin{equation}
y^{t}_k = \kappa y^{t-1}_k + (1 - \kappa) \left( \sum_j W^{\out}_{kj} z^{t}_j \right),
\label{eq:readout dyn}
\end{equation}
where $W^{\out}_{kj}$ denotes the synaptic weight from recurrent neuron $j$ to readout neuron $k$. The parameter $\kappa = e^{\Delta_t / \tau_{\out}}$ is the leak factor of the output neurons, with membrane time constant $\tau_{\out}$ set equal to the recurrent membrane time constant $\tau_m$ in all simulations.

Network performance is evaluated using a task-dependent loss function, defined in Eq.~\ref{eq:loss}:

\begin{equation}
E^{\task}(y, y^*) =
\begin{cases}
\frac{1}{2} \sum_{k,t} \left( y^{*,t}_k - y^t_k \right)^2, & \text{for regression tasks}, \\
-\sum_{k,t} \pi^{*,t}_k \log \pi^t_k, & \text{for classification tasks}.
\end{cases}
\label{eq:loss}
\end{equation}
Here, $y^{*,t}_k$ denotes the target value for regression tasks, while $\pi^{*,t}_k$ is the one-hot encoded target label for classification tasks. The predicted class probabilities $\pi^t_k$ are obtained by applying a softmax transformation to the readout activities, Eq.~\ref{eq:softmax}.
\begin{equation}
\pi^t_k = \frac{\exp(y^t_k)}{\sum_{k'} \exp(y^t_{k'})}.
\label{eq:softmax}
\end{equation}

\subsection{Connectivity and Spatial Embedding}
 All RSNNs employ fixed, random sparse connectivity for the external input and readout weights, and a fixed distance-dependent local connectivity pattern for the recurrent weights. This is implemented by defining a binary connectivity matrix $A^{\layer}_{ji}$ for each layer, where $A^{\layer}_{ji} = 1$ indicates the presence of a synaptic connection from presynaptic neuron $i$ to postsynaptic neuron $j$, and $A^{\layer}_{ji} = 0$ denotes the absence of a connection. Once initialized, each connectivity matrix $A^{\layer}_{ji}$ is kept fixed throughout training, and no structural plasticity or rewiring is allowed. Corresponding weights $W^{\layer}_{ji}$ with $A^{\layer}_{ji} = 0$ are set to zero and excluded from optimization.

The connectivity matrices $A^{\inn}_{ji}$ and $A^{\out}_{ji}$ for the external input and readout layers are initialized randomly, with $10\%$ of entries set to one. For the recurrent layer, the probability of a synaptic connection from neuron $i$ to neuron $j$ is defined as a function of the spatial distance $d_{ji}$ between the two neurons, with connection probability decreasing exponentially with the squared distance

\begin{equation}
P(A^{rec}_{ji} = 1) = \frac{1}{1 + e^{d_{ji}^2 / (4\sigma)}},
\label{eq:prob connection}
\end{equation}
where the parameter $\sigma$ controls the spatial decay of the connection probability. In all simulations, $\sigma$ is fixed to $0.012$, resulting in an average recurrent connectivity of approximately $10\%$.

To compute inter-neuronal distances, neurons in the recurrent layer are embedded in a two-dimensional grid with $h$ rows and $w$ columns, uniformly distributed over the unit square $\mathbb{I}^2 = [0,1] \times [0,1]$. Each neuron $i$, irrespective of whether it is an LIF or ALIF unit, is assigned a unique grid position $(x_i, y_i)$ without replacement, ensuring that each spatial location is occupied by at most one neuron. The distance $d_{ji}$ between neurons $i$ and $j$ is defined as the Euclidean distance under periodic boundary conditions:
\begin{equation}
d_{ji} = \sqrt{ \min(|x_i - x_j|, 1 - |x_i - x_j|)^2 + \min(|y_i - y_j|, 1 - |y_i - y_j|)^2 }.
\label{eq:distance}
\end{equation}

\section{E-prop}

Bellec et al.~\cite{bellec2020solution} introduced e-prop as an approximation to backpropagation through time (BPTT) for recurrent spiking neural networks, motivated by the goal of deriving a more biologically plausible learning rule. An equivalent learning principle for rate-based recurrent neural networks, termed random feedback local online learning (RFLO), was independently proposed by Murray~\cite{murray2019local}. The central idea underlying e-prop and RFLO, is to truncate the BPTT factorization, thereby enabling online and local weight updates. In this formulation, synaptic updates depend on three factors:  the activity of the pre- and postsynaptic neurons, and a neuron-specific feedback signal conveying information about the global network error. The complete derivation for e-prop updates is described in \cite{bellec2020solution}. Here, we briefly summarize its weight-update rules.

\subsection{Eligibility trace}
In e-prop, the eligibility trace associated with a synapse connecting presynaptic neuron $i$ to postsynaptic neuron $j$ is defined as the derivative of the postsynaptic neuron’s observable state, here, the spike output $z_j^t$, with respect to the synaptic weight $W_{ji}$. For adaptive leaky integrate-and-fire (ALIF) neurons, this definition leads to an eligibility trace that depends on multiple internal state variables. Specifically, the eligibility trace $e^t_{ji}$ can be decomposed into two components as
\begin{equation}
e^t_{ji} = \psi_j^t (e^t_{ji,v} - \beta_j e^t_{ji,a}),
\label{eq:eligibibility trace}
\end{equation}
where $e^t_{ji,v}$ corresponds to the contribution of the membrane potential variable $v_j$, given by   
\begin{equation}
e^t_{ji, v} = \mathcal{F}_{\alpha}(z^{t-1}_i),
\label{eq:v eligibibility trace}
\end{equation}
and $e^t_{ji,a}$, corresponds to the contribution of the adaptation variable $a_j$, given by
\begin{equation}
e^{t}_{ji, a} = \psi_j^{t-1} (1-\rho) e^{t-1}_{ji, v}  + (\rho - (1-\rho) \beta_j \psi_j^{t-1})e^{t-1}_{ji, a}.
\label{eq:a eligibibility trace}
\end{equation}
Here $\mathcal{F}_{\kappa}$ is a low-pass filter with exponential smoothing $\mathcal{F}_{\kappa}(x^t) = \kappa \mathcal{F}_{\kappa}(x^{t-1}) + (1 - \kappa)x^t$ and $ \psi_j^t$ is a surrogate gradient for $\frac{dz_j}{dv_j}$, defined as
\begin{equation}
\psi_j^t =
\begin{cases} 
0 & \text{during refractory period,}
\\
\frac{\gamma}{v_{\th}} \max(0, 1 - \frac{v^t_j - A^t_j}{v_{\th}})  & \text{otherwise} 
\end{cases}
\label{eq:surrogate gradient}
\end{equation}
with $\gamma$, a dampening factor, set to 0.3. The use of a surrogate gradient is necessary since the gradient of Heaviside step function is zero everywhere, except at zero, where it is not defined.

We notice that the same equations can be used by LIF neurons, by just setting $\beta_j=0$. The equations presented below describe the eligibility traces for recurrent synapses. The eligibility traces for external input synapses can be obtained analogously by replacing the presynaptic spike variable $z^{t-1}_i$ with the external input spike $x^t_i$.

\subsection{Credit Signal}
E-prop defines the credit signal $C_j^t$ (referred to as the \emph{learning signal} in \cite{bellec2020solution}) as the truncated derivative of the loss function $E$ with respect to the spike output $z_j^t$. The truncation ensures that the credit signal can be computed online and locally.

For regression tasks, the feedback credit signal $C^{t,\mathrm{direct}}_j$ is computed as
\begin{equation}
C^{t,\mathrm{direct}}_j = \sum_k W^{\out}_{kj} \left( y^{*,t}_k - y^t_k \right),
\label{eq:task_credit_signal}
\end{equation}
where $W^{\out}_{kj}$ denotes the readout weight connecting neuron $j$ to output unit $k$. We note that for neurons that are connected to the readout, this direct credit signal is absent (no feedback).

For classification tasks, the same expression applies by substituting the regression targets and outputs $y^{*,t}_k$ and $y^t_k$ with the corresponding class labels $\pi^{*,t}_k$ and predicted probabilities $\pi^t_k$, respectively.

\subsection{Update}
The updates for the recurrent and external input weights are calculated as a sum over time of the product of two key components: the credit signal $C^t_j$ (called the learning signal in the original paper) and the eligibility trace $e^t_{ji}$, 
\begin{equation}
\Delta W^{\text{rec/in}}_{ji} = \eta \sum_t{C^t_j \mathcal{F}_{\kappa}(e^t_{ji})},
\label{eq:e-prop}
\end{equation}
where $\eta$ is the learning rate and  $\mathcal{F}_{\kappa}$ is the low-pass filter. As shown in the supplemental materials of \cite{bellec2020solution}, the low-pass filtering of the eligibility trace is necessary for the online implementation of e-prop, due to the use of leaky readout neurons. We note that, to simplify notation,  Eq.~\ref{eq:e-prop} shows the update for the vanilla gradient descent optimizer, in which the learning rate is fixed. In our simulations, the updates are implemented with Adam \cite{kingma2014adam}.

In the setup with diffusion, we use $C^{t}_{j} = C^{t, direct}_j + C^{t, diff}_{j}$ as explained in Equation 1 in the main text.

The updates for the readout layers are computed as
\begin{equation}
\Delta W^{\out}_{kj} = \eta \sum_t{\left( y^{*,t}_k - y^t_k \right) \mathcal{F}_{\kappa}(z_j^t)}.
\label{eq:readout_update}
\end{equation}
\section{Firing rate regularization}
We include a firing-rate regularization term, following the approach described in the supplementary material of \cite{bellec2020solution}. This regularization penalizes deviations of the average firing rate of recurrent neurons from a target value, thereby encouraging stable and biologically plausible activity levels. The regularization loss is added directly to the task loss.
The firing-rate regularization loss is defined as
\begin{equation}
E^{reg} = \frac{c_{\reg}}{2} \sum_j \left( f^{\av}_j - f^{\target} \right)^2,
\label{eq:fire_reg}
\end{equation}
where $f^{\av}_j$ denotes the average firing rate of neuron $j$ computed across all trials in a batch, and $f^{\target}$ is the desired target firing rate. In all simulations, $f^{\target}$ is fixed to $10\,\Hz$. The adopted regularization coefficient $c_{\reg}$ differs across tasks, with specific values reported in Section~\ref{sec:params}.

The average firing rate $f^{\av}_j$ is computed as
\begin{equation}
f^{\av}_j = \frac{1}{B T \Delta t} \sum_{b=1}^{B} \sum_{t=1}^{T} z^{t,b}_j,
\label{eq:avg_firing_rate}
\end{equation}
where $B$ is the batch size, $T$ is the number of simulation time steps per trial, $\Delta t$ is the simulation time step, and $z^{t,b}_j$ denotes the spike output of neuron $j$ at time step $t$ in trial $b$.

Under the e-prop framework, this formulation leads to the update
\begin{equation}
\Delta W^{\mathrm{rec/in, reg}}_{ji} = \eta \, c_{\reg} \sum_t \frac{1}{B T \Delta t} \left( f_j^{\target} - f_j^{\av} \right) e_{ji}^t.
\label{eq:fire_reg_up}
\end{equation}
This contribution is then added to the task-dependent update. Importantly, since the firing-rate regularization term does not rely on a feedback signal from the output layer, its implementation is identical with or without diffusion. Moreover, because this term depends only on local neuronal activity, the resulting update is equivalent to the one obtained via BPTT for the corresponding component of the loss function.

\section{Tasks and Training}

\subsection{Tasks}
We evaluated the proposed learning rule by comparing its performance with BPTT and e-prop across the three benchmark tasks considered in \cite{liu2021cell}, with slight modifications. The three tasks are detailed below.

\subsubsection{Pattern Generation}
In the pattern generation task, the network is trained to reproduce a one-dimensional continuous target signal from spike-based input. The target signal is constructed as a weighted sum of five sinusoidal components. The network has to learn to generate the same target signal in response to eight distinct but fixed realizations of Poisson noise input. Using multiple fixed input realizations increases task difficulty.

For this task, we use an RSNN composed of $400$ LIF neurons with a membrane time constant $\tau_m = 30\,\ms$. The external input layer consists of $100$ neurons generating independent Poisson spike trains with a firing rate of $50\,\Hz$. The readout layer comprises a single leaky non-spiking unit with $\tau_{\out} = 20\,\ms$. The firing threshold is set to $v_{\th} = 0.03$.

Each trial lasts $2000\,\ms$. The target signal is generated as a weighted sum of five sinusoids with fixed frequencies of $0.5\,\Hz$, $1\,\Hz$, $2\,\Hz$, $3\,\Hz$, and $4\,\Hz$. For each run, the weights of the individual sinusoidal components are drawn randomly and normalized to sum to one. The resulting signal is then centered to have a zero mean.

\subsubsection{Delayed Match-to-Sample}
The delayed match-to-sample (DMS) task requires the network to classify whether two sequentially presented cues, separated by a temporal delay, belong to the same category. Each cue can take one of two values, $0$ or $1$. The network is required to output $1$ if both cues are identical (i.e., $1$--$1$ or $0$--$0$), and $0$ otherwise (i.e., $1$--$0$ or $0$--$1$).

We employ an RSNN composed of $50$ LIF neurons and $50$ ALIF neurons, with membrane and adaptation time constants $\tau_m = 20\,\ms$ and $\tau_{\adapt} = 1400\,\ms$, respectively. The adaptation strength parameter is set to $\beta = 1.8$, and the firing threshold is fixed to $v_{th} = 0.03$. The readout layer consists of two output neurons, one representing the probability of returning $1$ (matched cues) and the other $0$ (unmatched cues). Although binary classification can in principle be implemented using a single readout neuron, we retain two output units for consistency with \cite{bellec2020solution, liu2021cell}.

Each trial begins with a $50\,\ms$ fixation period, followed by the presentation of the first cue. Each cue is presented for $150\,\ms$, with a $700\,\ms$ delay separating the two cues. Immediately after the offset of the second cue, a $50\,\ms$ decision period is initiated. Cue values are drawn independently and randomly for each trial, with probability $p = 0.5$ of being $1$.

External input is provided by four distinct neuronal populations, each consisting of $20$ neurons and generating Poisson spike trains. Two populations encode the cue values, firing at $40\,\Hz$ when the corresponding cue equals $1$ and remaining silent otherwise. A third population encodes the decision period, firing at $40\,\Hz$ exclusively during this interval. Finally, a fourth population provides background activity, firing at $10\,\Hz$ throughout the entire trial to prevent the network from becoming quiescent during the delay period.

\subsubsection{Cue Accumulation}
In the cue accumulation task, a sequence of seven cues is presented sequentially, each separated by a brief temporal gap. Each cue appears on either the left or the right side. After the presentation of the final cue, a longer delay period follows, after which the agent must report on which side the majority of cues appeared, with $1$ indicating the right side and $0$ the left side.

We employ the same RSNN architecture as in the delayed match-to-sample task, consisting of $50$ LIF and $50$ ALIF neurons. The only modification concerns the adaptation time constant, which is set to $\tau_{adapt} = 2000\,\ms$ to accommodate the longer temporal integration demands of this task.

Each cue is presented for $100\,\ms$, followed by a $50\,\ms$ inter-cue interval. The side on which each cue appears is drawn independently and randomly for each cue and trial, with equal probability for both sides. After the final cue, a fixed delay period of $1000\,\ms$ is imposed, followed by a decision period lasting $150\,\ms$.

External input is provided by four neuronal populations, each consisting of $10$ neurons and generating Poisson spike trains. One population is activated whenever a cue appears on the left side, and another when a cue appears on the right side, both firing at a rate of $40\,\Hz$. In addition, a background noise population and a decision-time population are included, identical to those used in the delayed match-to-sample task.

\subsection{Training and Weight Updates}
For all tasks, the weights of each layer were initialized using Kaiming normal initialization \cite{he2015delving} and subsequently scaled by a layer-specific gain factor.

For the delayed match-to-sample and cue accumulation tasks, learning is performed using a cross-entropy loss, with target labels provided exclusively during the decision period. Consequently, credit signals are available only during this time window. Training is performed using batches of $64$ trials, with gradients averaged within each batch before applying weight updates.

For the pattern generation task, learning is based on a mean-squared error loss. Each batch consists of $8$ trials, where each trial receives a distinct realization of Poisson input noise; however, these input realizations are fixed across batches. Across all tasks, independently of the specific learning rule, parameter updates are performed using the Adam optimizer \cite{kingma2014adam}.

For the delayed match-to-sample and cue accumulation tasks, performance was evaluated on a test set comprising 512 trials generated independently of the training data.

For the pattern generation task, the Poisson input realizations were fixed throughout training and evaluation. Consequently, performance was assessed using the same input realizations as during training.

\section{Code Implementation}
All simulations are implemented in Python 3.12.3, using JAX library \cite{jax2018github} as the primary framework. The implementation will be released in a public repository upon publication.

\section{Simulation Parameters}
\label{sec:params}

\begin{table}[H]
\centering
\caption{Simulation parameters.}
\label{tab:task_parameters}
\begin{tabular}{lccc}
\toprule
\textbf{Parameter} 
& \textbf{Pattern Gen.} 
& \textbf{DMS} 
& \textbf{Cue Acc.} \\
\midrule
\multicolumn{4}{l}{\textbf{Architecture}} \\
\midrule

LIF & 400 & 50 & 50\\
ALIF & 0 & 50 & 50\\
Readout neurons & 1 & 2 & 2 \\
Grid shape & 20 x 20 & 10 x 10 & 10 x 10 \\
$\sigma$ & 0.012 & 0.012 & 0.012 \\
\midrule
\multicolumn{4}{l}{\textbf{Neuron Model}} \\
\midrule
$\tau_m$ (ms) & 30 & 20 & 20 \\
$\tau_{\out}$ (ms) & 30 & 20 & 20 \\
$\tau_{\adapt}$ (ms) & -- & 1400 & 2000 \\
$\beta$ (ALIF) & -- & 1.8 & 1.8 \\
$v_{\th}$ & 0.03 & 0.03 & 0.03 \\
Refractory period (ms) & 2 & 5 & 5\\
\midrule
\multicolumn{4}{l}{\textbf{Train}} \\
\midrule
Batch size & 8 & 64 & 64 \\
Task Loss function & MSE & Cross-entropy & Cross-entropy \\
$k$ (diffusion) & 0.75 & 0.75 & 0.75 \\
$\eta$  & 0.010 & 0.005  & 0.005 \\
$\gamma$ & 0.3 & 0.3  & 0.3 \\
$c_{\reg}$  & 0.010 & 0.010/0.100$^{*}$  & 0.005 \\
weight init gain (in, rec, out) & 1.0,1.0,1.0 & 0.5,0.1,0.5 & 1.0,1.0,1.0 \\
test set size  & 8 & 512  & 512 \\
\midrule
\multicolumn{4}{l}{\textbf{Others}} \\
\midrule
$\Delta t$ (ms) & 1 & 1 & 1 \\
\bottomrule
\end{tabular}
\vspace{1.5em} 
\footnotesize

$^{*}$ Two candidate values of $c_{\reg}$ were evaluated for both BPTT and e-prop (with and without diffusion). 
The higher value 0.100, yielded better learning performance for BPTT, whereas the lower value, 0.010, performed better for e-prop (with and without diffusion). 
Fig. 2 in the main text reports the learning curve under the best-performing value for each learning rule.

\end{table}

\begin{table}[H]
\centering
\caption{Pattern generation task parameters.}
\label{tab:pattern_task}
\begin{tabular}{ll}
\toprule
\multicolumn{2}{l}{\textbf{Temporal structure}} \\
\midrule
Trial duration & $2000\,\ms$ \\
Number of input realizations & $8$ (fixed across training) \\
Target frequencies & $0.5$, $1$, $2$, $3$, $4\,\Hz$ \\
Target normalization & Weights sum to $1$, zero-mean signal \\
\midrule
\multicolumn{2}{l}{\textbf{Input encoding}} \\
\midrule
Number of populations & $1$ \\
Neurons per population & $100$ \\
Firing rate & $50\,\Hz$ \\
\bottomrule
\end{tabular}
\end{table}

\begin{table}[H]
\centering
\caption{Delayed match-to-sample task parameters.}
\label{tab:dms_task}
\begin{tabular}{ll}
\toprule
\multicolumn{2}{l}{\textbf{Temporal structure}} \\
\midrule
Fixation period & $50\,\ms$ \\
Cue duration & $150\,\ms$ \\
Inter-cue delay & $700\,\ms$ \\
Decision period & $50\,\ms$ \\
Cue probability & $p=0.5$ \\
\midrule
\multicolumn{2}{l}{\textbf{Input encoding}} \\
\midrule
Number of populations & $4$ \\
Neurons per population & $20$ \\
Cue populations firing rate & $40\,\Hz$ (if cue = 1) \\
Decision population rate & $40\,\Hz$ \\
Background population rate & $10\,\Hz$ \\
\bottomrule
\end{tabular}
\end{table}

\begin{table}[H]
\centering
\caption{Cue accumulation task parameters.}
\label{tab:cueacc_task}
\begin{tabular}{ll}
\toprule
\multicolumn{2}{l}{\textbf{Temporal structure}} \\
\midrule
Number of cues & $7$ \\
Cue duration & $100\,\ms$ \\
Inter-cue interval & $50\,\ms$ \\
Post-cue delay & $1000\,\ms$ \\
Decision period & $150\,\ms$ \\
Cue side probability & $p=0.5$ \\
\midrule
\multicolumn{2}{l}{\textbf{Input encoding}} \\
\midrule
Number of populations & $4$ \\
Neurons per population & $10$ \\
Left/right population rate & $40\,\Hz$ \\
Decision population rate & $40\,\Hz$ \\
Background population rate & $10\,\Hz$ \\
\bottomrule
\end{tabular}
\end{table}

\end{document}